\documentstyle[12pt]{article}
\begin{document}

\begin{titlepage}

\begin{flushright}
OUTP-98-49P\\
hep-th/9806197\\
\end{flushright}

\begin{center}
\baselineskip 24pt
{\Large {\bf D-Instantons on the boundary}}\\
\vspace{.5cm}
\baselineskip 14pt 
{\large Ian I. Kogan}\\
kogan\,@\,thphys.ox.ac.uk \\
\vspace{.2cm}
{\large Gloria Luz\'on}\\
luzon\,@\,thphys.ox.ac.uk\\
\vspace{.2cm}
{\it Dept. of Physics, Theoretical Physics, University of Oxford,\\
  1 Keble Road, Oxford, OX1 3NP, United Kingdom}\\
\vspace{.2cm}
\end{center}

\begin{abstract}

The Maldacena's proposal has established an intriguing connection
between string theory in $AdS$ spaces and gauge theory. In this paper
we study the effects of adding $D(-1)$-branes to the system of
$D3$- or $(D1-D5)$-branes and we give arguments indicating that
$D(-1)$-branes are necessary to describe four and two dimensional instantons.
\end{abstract}

\end{titlepage}

\clearpage

\setcounter{section}{0}

\setcounter{equation}{0}
\def\theequation{\arabic{section}.\arabic{equation}}

\section{Introduction}

\
 
The description of supersymmetric gauge field theories by means of
superstrings has been a challenging problem. In brane theory, gauge
theory arises as an effective low energy description that is useful in
some regions in the moduli space of vacua \cite{Pol}. The underlying
dynamics is always the same - brane volume dynamics in string
theory. However, the physical interpretation of these objects has
remained quite obscure. Recently, based on previous works on the
structure of extremal black holes near the horizon \cite{Gi} and
absorption by these branes \cite{Gubs}, Maldacena has proposed an
exact correspondence between string theory on Anti-de-Sitter spaces (
times a compact space) and superconformal quantum field theories
living on the boundary of this space \cite{Gubs}. In particular, Type
IIB theory on  $AdS_5\times S^5$ should be dual to $N=4$ super
Yang-Mills on the boundary.  

This new AdS/CFT duality allows to describe many features of gauge
theory by means of branes. In this sense, perturbative computation of
correlation functions of local operator have been  presented in
\cite{GuKePo}\cite{WitHolo}, Wilson loops operators have ben computed
in \cite{MaldaWil} \cite{ReYe} and the description of baryons by
wrapping branes on the compact space has been given in \cite{WitBa}
and \cite{GO}. 

In this paper we are interested in finding branes configurations which
can lead to instantons solutions of $d$  Yang-Mills field theory,
the $d$ dimensional space being the $AdS_{d+1}$ boundary.
 If we construct systems of
branes within the configuration of branes wich causes the $AdS$
background, 
in the AdS/CFT duality picture,
the `small' branes should behave as physical objects for the gauge
field theory on the boundary. The natural $D$-brane objects to be
identified with boundary instantons are the $D$-Instantons (
$D(-1)$-branes). The world volume of these tiny objects is just a
point. Therefore, as we wil see, they hardly alter the basic structure of the space.

In section 2  of the paper we will remind that the $AdS_5\times S^5$ space
appears as a supergravity solution for $D3$-branes in the low energy
limit. Then, in view of the fundamental role of
supergravity solutions of branes near the horizon, looking for
solutions of composite branes ( branes within branes) seems to  be
important.  The description of smeared solutions of branes ending on
branes  have been extensively treated in literature \cite{Papa} and
some localized solutions have been also found for specific
configurations of 
branes \cite{Tse}. 
 
In section 3, we will focus on  solutions of the $D-$Instanton
within the $N$ $D3$-branes. We are interested in the configuration on
the boundary and, therefore, in localized solutions. Here we will
characterize the $D-$Instanton by means of a quasi conformally
invariant harmonic function connected to the dilaton field and we will
show how the presence of $D-$Instantons affects the structure of the
space by developping throats which connect different vacua. 

In section 4, we will describe the field theory on the branes, the
coupling to the gravitational fields and the new features,
self-duality and non-trivial instanton number, which the existence of
$D(-1)$-branes induces on the gauge fields.

In section 5, we will approach the supergravity solution for
$D$-Instantons in the bulk to the boundary. Here we will find that the
system of $D(-1)$-branes stuck to the $N$ $D3$-branes  behaves as
small $SU(2)$ Yang-Mills instantons. We will obtain  the Yang-Mills
solution corresponding to the system, the moduli space of both of them
and the instanton  mesure

Finally, in section 6,  we will discuss supergravity
solutions with lower supersymmetry ( $D(-1)$- branes within a
$D1$-$D5$ system) and its relation to $d=2$ instantons.

We will conclude with a brief discussion of the results and
new possibilities for a further research.

While this paper was being prepared for publication, we learned about a recent work
on the same subject \cite{ChuHoWu}.

\setcounter{equation}{0}

\section{The $AdS_5\times S^5$ background}

\ 

D-branes can be identified with BPS-saturated, R-R charged $p-$brane
solutions. The Supergravity solution carrying D$p$ charge  can be
characterized in terms of a function $H_p (x_{_\perp})$ which is harmonic with
respect to the directions transversal to the world volume
\cite{HoStr}; i.e.  it verifies the $10-(p+1)$ dimensional Laplace equation
\begin{equation}
\label{laplacian}
 \partial_{x_{_\perp}}^2 H_p(x_{_\perp})=0.
\end{equation}
Asuming that $H_p$ depends only on the radial coordinates
$r=\sqrt{x_{p+1}^2+ ... +x_9^2}$, we can solve (\ref{laplacian}) to get
\begin{equation}
\label{hp}
H_p(r)= 1+{ Q_p \over r^{(7-p)}}.
\end{equation}
where the charge $Q_p$ is related to the string tension $T\equiv
(2\pi\alpha)^{-1}$ 
\begin{equation}
\label{charge}
Q_p=g(
2\pi)^{(5-p)/2}(2\pi\alpha^\prime)^{(7-p)/2}(2\pi^{(7-p)/2}/\Gamma((7-p)/2))^{-1}.
\end{equation}
Then, the euclidean metric in string frame is
\begin{equation}
\label{pmetric}
  ds_p^2= H_p^{-1/2}( dx_0^2+...+ dx_p^2)+ H_p^{1/2}( dx_{(p+1)}^2+...+
				dx_9^2)
\end{equation}
with the dilaton field $\phi$  given by
\begin{equation}
\label{pdilaton}
e^{2\phi}=H_p^{(3-p)/2}.
\end{equation}
The R-R gauge field strength associated with the $p-$brane can be also
expressed in terms of the harmonic function \cite{BerRoo}
\begin{equation}
\label{fp1}
F^{(p+2)}=d{1\over H_p}\wedge dx^0.....\wedge dx^p, 
\end{equation} 
in case $p\le 3$ ( they carry electric charge) and its dual for $p\ge 3$            
( they carry magnetic charge). Notice that the case $p=3$ is
self-dual. 
 The flux of the dual   field strengths  on theirs $S^{(8-p)}$
transversal spheres fix the value of the charges.

Now we consider the string background describing one $3-$brane
\begin{equation}
\label{3metric}
  ds_3^2= H_3^{-1/2} d\vec x^2+ H_3^{1/2}( dr^2+ r^2 d\Omega_5^2),
  \ \ \ H_3= 1+ {4\pi g \alpha^{\prime 2}\over r^4}
\end{equation}
where $\vec x =( x_0,..., x_3)$ denotes the four dimensional world
volume of the $3-$brane and $d\Omega_5^2$ is the metric in $S^5$. There
is no dilaton field\footnote{That corresponds to the fact that the
theory on $D3$-branes is conformal.}, then the string frame and the
Einstein frame are identical
\begin{equation}
ds_E^2=ds_s^2.
\end{equation}
In case of $N$ parallel $3-$branes, the BPS condition of  `No force' implies 
\begin{equation}
\label{h3}
H_3= 1+ {4N\pi g \alpha^{\prime 2}\over r^4 }.
\end{equation}

In this paper we will consider the low energy effective theory. In this
limit 
\begin{equation}
\label{limit}
\alpha^\prime \rightarrow 0, \ \ \ U\equiv {r\over \alpha^\prime} = \rm{fixed}
\end{equation}
the field theory on the $3-$brane decouples from the bulk
\cite{Malda}. In this limit, the constant term in the harmonic function 
(\ref{h3}) can be neglected and the metric, in terms of the new variable
$U$, becomes
\begin{equation}
 ds_3^2= \alpha^\prime\left[{u^2\over\sqrt{4\pi g N}} d\vec x^2+ \sqrt{4\pi g N}
  {du^2\over u^2}+ \sqrt{4\pi g N} d\Omega_5^2 \right]
\end{equation}
which describes the $AdS_5\times S^5$ space and remains constant in 
$\alpha^\prime$ units.
 Notice that $S^5$ and
$AdS_5$ have the same radius and, being spaces of opposite curvature, 
the total scalar curvature of the $AdS_5\times S^5$ space is identically zero.

The supersymmetry group of euclidean $AdS_5\times S^5$, $SO(1,5)\times
SO(6)$, and, as it has been shown in \cite{WitHolo}, the conformal
compactification of $AdS_5$, on which $SO(1,5)$ acts, is the sphere
$S^4$.
The supersymmetric group is the same as the superconformal group in
four dimensions \cite{HaLoSo}. This fact led to Maldacena's proposal 
\cite{Malda}. According to it, when the effective coupling $g_e=gN$
becomes large , the $N=4$ superconformal  theory on the boundary is governed by
supergravity on   $AdS_5\times S^5$ where perturbation theory be can trusted.

Rescaling  the $u\rightarrow u \lambda^{-1}$ and $ \vec x\rightarrow
\lambda \vec x$ variables by the factor $\lambda^4= 4\pi gN$, we obtain the
metric 
\begin{equation}
\label{umetric}
ds_3^2= \alpha^\prime \sqrt{4\pi g N}\left[u^2 d\vec x^2+ 
  {du^2\over u^2}+  d\Omega_5^2 \right]
\end{equation}
and, using the inverse variable $z=1/u$, the metric
\begin{equation}
\label{zmetric}
ds_3^2= \alpha^\prime \sqrt{4\pi g N}\left[{1\over z^2}( d\vec x^2+ 
  dz^2)+  d\Omega_5^2 \right]
\end{equation}
which we will use, ignoring constant factors, from now on.
In this representation of $AdS_5$ as the upper space $z>0$, the boundary consist of a copy
of $R^4\times S^5$, at $z=0$, and a single point at $z=\infty$ which
compactifies $R^5\times S^5$ to $S^4\times S^5$.

\setcounter{equation}{0}

\section{D-Instantons in $AdS_5\times S^5$}
\

The $p-(p+4)$ system of branes is a BPS bound state which preserves
$1/4$ of the supersymmetries; i.e, we are dealing with a $N=2$ SUSY
theory. These  $p-(p+4)$ systems are marginally bound. This means that
the total energy is the sum of the energies.

We are interested in placing D-Instantons in the  $AdS_5\times S^5$
space. Therefore, we will  construct supergravity solutions of
$D(-1)$-branes within a collection of $N$  $D3-$branes in the
decouplig limit. The solutions are required to   preserve as much
as possible the symmetries of this space. The
localized  solution should correspond to localized instantons in the  
four dimensional theory.

The supergravity background for such a system is represented by the
following metric
\begin{equation}
\label{1zmetric}
ds_{(-1,3)}^2= H_{-1}^{1/2}\left[{1\over z^2}( d\vec x^2+ 
  dz^2)+  d\Omega_5^2 \right],
\end{equation}
where we have taken off the prefactor which appears in
(\ref{zmetric}), the $1-$ and $4-$forms
\begin{equation}
\begin{array}{rl}
 F^{(1)}&= dH_{-1}^{-1} \nonumber \\ 
 F^{(4)}&= d({1\over z^4}) (dx_0^2\wedge...\wedge dx_3^2),
\label{gsfields}
\end{array}
\end{equation}
and dilaton field
\begin{equation}
\label{ddilaton}
e^\phi=H_{-1}.
\end{equation}
In this case the string and the Einstein  metric do not coincide
\begin{equation}
\label{smet}
ds_{string}^2= e^{\phi/2}ds_E
\end{equation}
and, due to the relation (\ref{ddilaton}), the Einstein metric still
corresponds to $AdS_5\times S^5$ space.

Assuming that the harmonic function $H_{-1}$ depends only on the
coordinates of the $D3-$brane world volume  $\vec
x= (x_0,x_1,x_2,x_3)$ and on $z$, it must satisfy the
Laplace equation in the ten-dimensional curved transverse space
{\footnote{ This condition  preserves  the flatness of the space.}}
\begin{equation}
\label{l13}
\left[ \triangle_{||}+ z^3{\partial \ \over \partial z} {1\over z^3}{\partial \
\over \partial z}
\right] H_{-1}(\vec x,z)=0
\end{equation}
where $\triangle_{||}$ represents the laplacian in the four
dimensional space.
This condition is invariant under translations in $x_i,
n=0,..3$ and under the $SO(4)\times SO(6)$ symmetry group of
$SO(1,5)$. However, a conformal transformation that maps the point at infinity to the origin
\begin{equation}
\label{ctrans}
z\leftrightarrow {z\over z^2+x^2} \ \ \ \  x_i\leftrightarrow {x_i\over
z^2+x^2}, \ i=0, ...,3
\end{equation}
does change the harmonic function leaving invariant the
laplacian. Therefore, for a given solution $H_{-1}(\vec x,z)$ of (\ref{l13}),
its transformed function under (\ref{ctrans}), $H_{-1}^t=H_{-1}({x\over z^2 +
x^2},{x\over z^2 + x^2})$ is also a solution.
Returning to the metric (\ref{1zmetric}),  it is straight to see that
it exhibits this same behaviour; i.e., it is invariant under
$SO(4)\times SO(6)$ and only the $H_{-1}^{-1/2}$ prefactor changes under
(\ref{ctrans}).

In the representation of the bulk as the upper space $z>0$ we have
been using, the transformation (\ref{ctrans}) interchanges the two
boundary regions. Now at infinity ($z^t=0$) we have a copy of $R^4$
and the boundary at $z=0$ ($z^t=\rm{infinity}$) is just a point. Then, the
compactified space is still $S^4\times S^5$, but the normal vector has
flipped. This change of orientation transforms the D-instanton into
the anti D-Instanton \cite{CaMal} and changes the sign of the flux of the $1-$form
defined in (\ref{gsfields}) on $S^4\times S^5$. It will be clear later
when we relate $D(-1)$-branes to Yang-Mills instantons that this
operation precisely corresponds to 
coordinate inversion which  sends the pseudoparticle with $q=Q$ into the
antiparticle with $q=-Q$ \cite{JaRe}.

Now  we will construct specific supergravity solutions for
the $D-$Instanton sitting on the $D3$-branes. That means $D(-1)-$brane 
configuration centered at the $u_0=0$.

A solution of (\ref{l13}) singular at a point on  the boundary at $z=0$ can be shown to be
\begin{equation}
\label{sc2}
e^{\phi}= { c z^4\over ( (\vec x-\vec x_{(0)})^2+z^2)^4}
\end{equation}
where the constant $c$ is related to the charge of the $D-$Instanton
and $ \vec x_{(0)}$ is the position in $R^4$.
 Its transformed  function under (\ref{ctrans})  will give
us the solution singular at infinity
\begin{equation}
\label{sc1}
e^{\phi}=  c z^4.
\end{equation}

Then, the conformal transformation (\ref{ctrans}), transforms
(\ref{sc2}) into (\ref{sc1}), leaving invariant the underlying
$AdS_5\times S^5$ metric. This fact allows us relate the behaviour of
the system at the origin and at infinity.
 In the string frame (\ref{smet}), the configuration consists
of two asymptotic $AdS_5\times S^5$ spaces, one at the singularity at
the origin and the other at infinity, which are connected by a
throat \cite{GiGrPe}\cite{BerBehr}.The space is geodesically complete, so in this sense is not
singular. That is analogus to classical instantons in field theories
which join two different vacua of the theory.

The constant $c$ are related to the electric charge $Q^{-1}$ of the
$D-$Instanton. This can be calculated as the flux of the $9-$form,
\begin{equation}
\label{dualf}
F^{9}= e^{2\phi(x,z)}*dH_{-1}^{-1}(x,z)=-e^{\phi(x,z)}*d\phi(x,z),
\end{equation}
 dual to $F^{1}$, on the $S^4\times S^5$ space
$$
 Q^{-1}=  \frac{1}{Vol(\Sigma)Vol(S_5)}\int_{AdS_5\times S^5}*d*
d(e^{\phi}) 
$$
\begin{equation} 
       =  {c\over  Vol(\Sigma)} 
\int_x d^4x\int_0^\infty dz\ \partial_z {1\over z^3}
 \partial_z{ z^4\over(x^2+z^2)^4 }  
\end{equation}
$$= {c\over 4 Vol(\Sigma)}~
\lim_{z\rightarrow 0}\int_x d^4x  {1\over z^4}
 { z^4\over(x^2+z^2)^4 }= {c\over 4}
$$
where $\Sigma=\partial AdS_5$.
A quantization condition relates the R-R electric charges of
$p-$branes to magnetic charges of
$(p+6)$-branes\cite{Nepo}\cite{Tei}. The associated R-R field
strength $F^{(9)}$ is nine-dimensional and, therefore, the flux of its
dual has to be calculated on the $S^1$ sphere. For this reason its
charge is quantized and that gives a quantized charge for the
$D(-1)-$brane. In hte following we will take $c=1$ for  one
$D-$Instanton. And, as a BPS state,  for $n$ instantons we will
have  $c=n$.

Due to the linearity of the laplacian in (\ref{l13})  the
multiinstanton solution
 will be a superposition of solutions 
\begin{equation}
\label{sc2m}
e^{\phi}= \sum_i{ c_i \  z^4\over ( (\vec x-\vec x^i_{(0)})^2+z^2)^4}
\end{equation}
as corresponding to a BPS state. At every of these singularities the
space will develop a throat.

Another set of solutions of (\ref{l13}) can be found by factorizing
$$H_{-1}(x,z)=F(x)G(z)$$ as a product of two functions. Then, we have
\begin{equation}
\label{sc4}
e^{\phi}= c {z^4 \over (x^2+ z^2)^4}{ (x^2+ z^2)^2 \over x^2}
\end{equation}
singular at the origin,
and its transformed under (\ref{ctrans}) function
\begin{equation}
\label{sc3}
e^{\phi}= c  z^4 {1 \over x^2},
\end{equation}
singular at the infinity. 
When gravity decouples from field theory, all the gravitational fields
should behave as Green's functions. Note that these last solutions are
more regular on the boundary and they may not be a good description
for the dilaton field.

Finally, we could also consider supergravity solutions for the
$D(-1)$-brane smeared (as opposite to localized) in the $D3$-branes
world volume. That can be achieved by integrating  the $x_i, (
i=0,...,3)$ coordinates ( using the $AdS_5\times S^5$) metric to obtain
\begin{equation}
\label{smso}
e^\phi= z^4,
\end{equation}
but this solution does tell us too much about the structure of the
space because the transformation (\ref{ctrans}) does not act longer.

\section{Field Theory on the branes}

\

We will describe in this section the action on the branes
($AdS_5\times S^5$)  to wich the superconformal field theory on the
boundary is dual. 

The bosonic part of the effective low energy field theory for a $D3$-brane
brane, in the R-R sector, is given by the Born-Infeld action
\begin{equation}
\label{bi}
S= \int_{4} \ \ e^{-\phi} \sqrt{ det (G + F)}
\end{equation}
where $G$ is the metric on the brane world-volume ( which , in static
gauge,  includes also  six scalar fields  ) and $F$ is the gauge
field strength. From this action is clear  the coupling 
\begin{equation}
\label{cym}
C_1= {1\over 4}\int_4 e^{-\phi}F^2.
\end{equation}
The presence of a $D(-1)$-brane will induces a coupling of the gauge
field to the axion field $A^{(0)}$, $F^{(1)}=dA^{(0)} $, which arises
from a Chern-Simons term
\begin{equation}
\label{csa}
C_2=-{1\over 4} \int_4 A^{(0)} F\wedge F.
\end{equation}
The fact that the instanton number ${1\over 8 \pi^2}\int_4 A^{(0)}
F\wedge F$
 carries $(p-4)$-brane charge was observed by Witten \cite{WittenSm}
for $5-$ and $9-$branes and as a general result by Douglas \cite{Douglas}.
Then, the presence of a gauge field with non-trivial instanton number
is necessary in order to induce the $D(-1)$-brane charge on a
$D3$-brane. Moreover, we know that the $D(-1)$-$D3$ system is a BPS
marginal configuration and, for correspondence to the properties of
such a configuration, we must require the self-duality condition on this
field. Therefore,  the $D(-1)$-$D3$ action can be described as that of
the $D3$-brane with an extra self dual gauge field \cite{CheTse}.

From the relation $F^{(1)}= d e^{-\phi}$ between the gravitational
filed strenght and the dilaton, it follows that the couplings in
(\ref{cym}) and (\ref{csa}) are similar. Then, in the limit where
the field theory on the brane decouples from the bulk, the properties
of self-duality and non trivial instanton number will remain.

\section{$D-$Instantons localized on the boundary and small Instantons}

\

 The $AdS/CFT$ correspondence
 tells
us that in the limit where gravity decouples, the theory on the branes
should be dual to the super Yang-Mills theory on the boundary of
$AdS_5\times S^5$. Therefore, the system of $D-$Instantons sitting on
$D3$-branes should  describe $YM$ instantons on the boundary. 
Here we have a configuration space of Super Yang-Mills
 gauge fields and, as we have discussed in the preceding section,
among them there exist self-dual configuration with non-trivial
instanton number; i.e., Yang-Mills instantons. Then, by studing  the
new gravitational fields introduced by the $D-$Instantons we will try
to find out some properties of Yang-Mills instantons. In this sense,
both configurations are dual.

In the following we will focus on the first set of supergravity
solutions (\ref{sc2}),(\ref{sc1}) and (\ref{sc2m}) for one instanton , 
though some of the arguments could be extended to the second one.
 We can see that the solution approaches on the boundary to
\begin{equation}
\label{sc2b}
e^{\phi}= {  z^4\over ( (\vec x-\vec x_{(0)})^2+z^2)^4} , \ \ z\rightarrow 0.
\end{equation}
 This kind of
singularities  in the dilaton have been related to
Yang-Mills instanton in the limit of small scale size \cite{CaHaSt} in
the context of heterotic strings.
The  parameters which describe our D-Instanton solution (\ref{sc2b})
on the boundary are its position given by the four coordinates $\vec
x_{(0)}$,  and the
parameter $z$ which can be understood as an $UV$ regularization of the
Yang-Mills boundary theory \cite{SuWit}. Then, we can identify the
position of the $D-$Instanton with that of the Yang-Mills instanton on
the boundary and the regulator $z$  with the size of the small
instanton \cite{CaHaSt}. 
 Note that we have placed
 the $D-$Instanton at the point where the collection of $D3-$branes
were sitting originally $u_0=0$ and, interpreting $u$ as a scale of
energies, that means a point in the IR. So, we can see that, in the
holographic spirit, infrared effects in the bulk theory have been
reflected as ultraviolet effects on the boundary theory.

Following with the identifications, the
electric R-R charge carried by the $D-$Instanton which flow through
the throat might represent in the dual picture the instanton
number. Working on a regulated boundary \cite{MaldaWil}-\cite{WitBa}
  and from
\begin{eqnarray}
\label{inst}
n\sim Q^{-1} \sim \lim_{z\rightarrow 0} \int_x d^4x  {1\over z^4}
 { z^4\over(x^2+z^2)^4 },
\end{eqnarray}
where we have ignore constant factors, we obtain
\begin{equation}
\label{fins}
Tr\{F\wedge F\} \sim  {z^4\over(x^2+z^2)^4 }
\end{equation}
which is the correct expression for Yang-Mills instantons of size $z$. Then,
thinking of gauge fiels $A_i$ on $S^3\subset S^4$ we arrive to the known
pseudoparticle solution 
\begin{equation}
\label{ginst}
A_i= {z^2x^2\over (x^2+z^2)} g^{-1}\partial_i g,
\end{equation}
where $g=(x_3-i \vec x\vec \sigma)/(x_ix^i)^{1/2}\ $ is the imbedding of
$S^3$ into the group manifold of $SU(2)$. We see that the size of this
instanton shrinks to zero as the regulator parameter $z$ goes to zero.

We will discuss next the moduli space of $D-$Instantons in
order to compare it to that of $SU(2)$ instantons on $S^4$. 

 We have characterized the $D-$Instanton by the
harmonic function $H_{-1}(\vec x, z)$ solution of the Laplacian equation
(\ref{l13}) in $AdS_5\times S^5$ space. The solutions we have found
are invariant under rotation $SO(6)$ in  $S^5$ and rotations $SO(5)$
in $S^4$, but there exist a $SO(1,5)$ (\ref{ctrans}) transformation  which transforms
it into other. Then, the moduli space of our supergravity solution is
\begin{equation}
\label{moduli}
{SO(1,5)\over SO(5)} \times { SO(6)\over SO(6)} 
\end{equation}
which coincides with the moduli space of $SU(2)$ instantons on $S^4$ of
instanton number one \cite{JaNoRe}. Let us note that $ AdS_5=SO(1,5)/
SO(5)$. Then, the moduli space of one instanton solution is $AdS_5$
\cite{WitBa} and coordinates $\vec x, z \ $ in $AdS_5$ are coordinates
of instanton moduli space. Therefore, the natural measure on this
moduli space is $d\mu=\sqrt{g} dx_0...dx_4 dz$ in $AdS_5$ space. Using the
$AdS_5$ metric
\begin{equation}
\label{admetric}
ds^2={1\over z^2} (dz^2+ d\vec x^2),
\end{equation}
 the measure on the moduli space can be expressed as
\begin{equation}
\label{measure}
d\mu= {d^4x \ dz\over z^5}
\end{equation}
which is the well known instanton measure \cite{Poly}.

Let us consider now $M$ $D-$Instantons. The moduli space
of  instanton with charge $M$ and  gauge group $SU(N)$ has different
components. Each of this components describes how the $M$ instantons
have been placed in the $SU(2)$ factors of $SU(N)$. That seems closely
related, in the dual picture of $D-$Instantons, to the way in which
the $M$ $D(-1)$-branes have been attached to the $N$
$D3$-branes. Then, as an example, the symmetric component of the moduli space of $M$
instantons would correspond to the $M$ $D(-1)$-branes stuck to
different $M$ $D3$-branes (symmetrized).
\newpage
\section{ Topological defects in two dimensions}
\

Let us describe now another system of branes where the
presence of $D$-Instantons within leads to topological defects in the
$d=2$ gauge field on the boundary. We will consider a collection of $N_5$
$D5$-branes with world volume coordinates $x_i$, $(i=0,1,..,5)$
wrapping on a compact manifold $M_4$ and
$N_1$ $D$-strings parallel to  the first collection in $x_0,x_1$
coordinates and smeared in the $M_4$ coordinates. All of them are
sitting at the point $x_6=x_7=x_8=x_9=0 \ $ of the transverse space. As a BPS bound
state, the number of supersymmetries broken by such a state is $1/4$. 

The exact string background describing this configuration is
represented by the conformal sigma-model with the following metric
\begin{equation}
\label{met15}
\begin{array}{rl}
ds^2=
&(H_1H_5)^{-1/2}(dx_0^2+dx_1^2)+H_1^{1/2}H_5^{-1/2}(dx_2^2+...+dx_5^2)
\\ \nonumber + &
 (H_1H_5)^{1/2}(dx_6^2+...+dx_9^2)
\end{array}
\end{equation}
where the harmonic functions $H_1$ and $H_5$, depending only on the
transverse radial coordinate $r=\sqrt{x_6^2+...+x_9^2}$, are 
\begin{equation}
\label{hh15}
H_1= 1+ {g\alpha^\prime N_1\over v r^2}, \ \ \ \ H_5= 1+
{g\alpha^\prime N_5\over  r^2},
\end{equation}
 non-trivial seven and three strength field forms (\ref{fp1}) and
dilaton field
\begin{equation}
\label{d15}
e^{2 \phi}= H_1/H_5 .
\end{equation}

In the decoupling limit \cite{Malda}
\begin{equation}
\label{de15}
\alpha^\prime\rightarrow 0, \ \ \ u= {r\over \alpha^\prime}={\rm fixed},
\ \ \ v= {V_4\over(2\pi)^4 \alpha^{\prime 2}}={\rm fixed}, \ \ \
g_6={g\over \sqrt{v}}
\end{equation}
with $V_4$ being the $M_4$ volume, we find the low energy metric
\begin{equation}
\begin{array}{rl}
\label{lmet15}
ds^2= \alpha^\prime g_6 \sqrt{N_1N_5} [ & u^2 (dx_0^2+dx_1^2)+{du^2\over u^2}+
d\Omega_3^2 \\ \nonumber + &
\beta(N_1,N_5)(dx_2^2+...+dx_5^2)],
\end{array}
\end{equation}
where we have used the rescaling of section 2  with
$\lambda^4= g_6^2 N_1 N_5 \ $, $d\Omega_3^2$ is the metric in the unit
three sphere
 and  $\beta(N_1,N_5) = (\alpha^\prime g_6 N_5
v^{1/2})^{-1}$. This metric
describes the $AdS_3 \times S^3 \times M_4$ space, where the $M_4$
factor is a compact hiperk\"aler manifold ( $T^4$ or $K^3$ ) depending on the charges of the branes. Note
that in this limit the dilaton field is a constant $e^{2\phi}=N_1/(N_5
v)$. Any of our further results will not depend on the moduli space
$M_4$ or on constant factors of (\ref{lmet15}), then we will consider the seven
dimensional metric, wich in terms of the inverse variable $z=1/u$
reads
\begin{equation}
\label{lzmet15}
ds^2= \left[  {1\over z^2} (dx_0^2+dx_1^2+ dz^2)+
d\Omega_3^2 \right] .
\end{equation}
In this representation, the $AdS_3\times S^3$ space is the upper space $z>0$,
its supersymmetric group is $SO(1,3)\times SO(4)$ and the conformal
compactified  boundary
consist of the plane $R^2$ at $z=0$ and a point at infinity.
The AdS/CFT picture tell us that the type $IIB$ string theory on
$AdS_3\times S^3$ in the limit of large $N$ is dual to the $N=2$
superconformal Yang-Mills theory on the boundary.

Now, as  in section 3, we can add $D$-Instantons to the system
of branes. This new collection of branes  breaks $1/2$ of the
supersymmetries. In this case the $p-(p+2)$ BPS system of branes is
truly bound. The supergravity solution for the system is
\begin{equation}
\label{15zmetric}
ds^2= H_{-1}^{1/2}\left[{1\over z^2}( dx_0^2+ dx_1^2+
  dz^2)+  d\Omega_3^2 \right],
\end{equation}
with $ H_{-1}$ satisfying the laplace condiction
\begin{equation}
\label{l15}
\left[ \triangle_{||}+ z{\partial \ \over \partial z} {1\over z}{\partial \
\over \partial z}
\right] H_{-1}(\vec x,z)=0,
\end{equation}
where $\triangle_{||}$ represents the laplacian in the plane. We have
also the $F^{(1)}$ strength given in (\ref{gsfields}) and the dilaton
field of (\ref{ddilaton}) which leaves the  metric in the Einstein
frame  invariant.

Following the arguments of section 3, we find the solution 
\begin{equation}
\label{h15}
e^{\phi}= c {z^2\over ((\vec x-\vec x_{(0)})^2+ z^2)^2},
\end{equation}
invariant under $SO(3)\times SO(4)$. Its transformed function under  (\ref{ctrans})
 is the solution singular at infinity
\begin{equation}
\label{h15i}
e^{\phi}= c z^2.
\end{equation}
In this case, the $D-$Instanton connects two $AdS_3\times S^3$ spaces
by a throat located at $(\vec x_{(0)}, z=0 )$ and at
infinity. The charge which flows through the throat is given by
$$
 Q^{-1}=  {1\over Vol(\Sigma)Vol(S_3)}\int_{AdS_3\times S^3}*d*
d(e^{\phi})
$$ 
\begin{equation}
       =  {c\over  Vol(\Sigma)} \int_x d^2x\int_0^\infty dz\
\partial_z
 {1\over z}
 \partial_z{ z^2\over(x^2+z^2)^2 } 
\end{equation}
$$
=  {c\over 2 Vol(\Sigma)}
~\lim_{z\rightarrow 0}\int_x d^2x  {1\over z^2}
 { z^2\over(x^2+z^2)^2 }= {c\over 2}.
$$
With regard to the field theory on the branes, the presence of
$D(-1)$-branes induces a non trivial monopole (or two dimensional
instanton) number
\begin{equation}
-{1\over 2}\int_2 A^{(0)} F
\end{equation}
which couples to the axion field.

Now, after having described the supergravity solution, we are able to
repeat the discussion of section 5 in order to indentify the solution
on the boundary
\begin{equation}
\label{sc2b2}
e^{\phi}= {  z^2\over ( (\vec x-\vec x_{(0)})^2+z^2)^2} , \ \ z\rightarrow 0.
\end{equation}
with two dimensional gauge instantons {\footnote{ A detailed
discussion about gauge field description of $2d$ instantons can be
found in Polyakov's book \cite{Poly} or in NSVZ's review \cite{NSVZ}.}}.

Here we find again the same kind of singularities in the dilaton field
related to small instantons. The parameter $z$ plays the role of an
$UV$ regulator for the instanton size.
This is another example of the $IR$-$UV$ connection. The infrared
regulator $1/z_0=u_0=0$ we have used to place the $D(-1)$-brane in the
bulk has been translated to an $UV$ regulator on the boundary.

The correct expression for the two dimensional instantons can be
obtained from the identification of the monopole number with the
charge of the $D-$Instanton
\begin{equation}
\label{mon}
\int_x d^2 x \  F \sim \lim_{z\rightarrow 0}\int_x d^2 x \ {1\over z^2}{z^2\over
(x^2+z^2)^2}
\end{equation}
which leads to the two dimensional gauge field solution
\begin{equation}
\label{gmon}
A_{\bar w}= { z^2 \vert w\vert^2 \over (\vert w\vert^2 + z^2)^2} g^{-1}\partial_{\bar w}
g ,\ \ \ A_w= -\bar{ A_{\bar w}},
\end{equation}
$w, \bar w$ being complex coordinates on the plane and 
\begin{equation}
\nonumber
 g={1\over 1+ \vert w\vert^2} \left( \begin{array}{cc} 1 & -w \\ \bar w & 1 \end{array}
\right)
\end{equation}
 the embedding of the plane into the group manifold of
$SU(2)$.

Finally, we observe that the moduli space of one $D$-Instanton in
$AdS_3\times S^3$,
characterized by the harmonic function $H_{-1}$,
\begin{equation}
\label{modmo}
{SO(1,3)\over SO(3)}\times {SO(4)\over SO(4)}
\end{equation}
agrees with that of the two dimensional instantons on $S^2$ of $\ 2d\ $
instanton number one. Again, the coordinates of the $AdS$ space 
$AdS_3= SO(1,3)/SO(3)$ are the coordinates of the instanton moduli
space. Then, from the $AdS_3$ metric
\begin{equation}
\label{admetric3}
ds^2={1\over z^2} (dz^2+ d\vec x_0^2+ dx_1^2),
\end{equation}
we obtain the measure on the instanton moduli space
\begin{equation}
\label{measure3}
d\mu= {d^2x \ dz\over z^3}.
\end{equation}
It is easy to see that this measure is a right $\ 2d$ instanton measure
by analizing the simplest example of instanton in $CP( O(3))$
model \cite{Poly}. This instanton is given by analytic functions
${z-a\over z-b}$ and its moduli space by the two complex coordinates $a$ and $b$. The
one-instanton contribution is required to be dimensionless and, by
translation invariance, to depend on $\vert a-b\vert $. Therefore, it
must be 
\begin{equation}
\label{meas}
d\mu={ d^2 a d^2 b\over\vert a-b\vert^4}= {d^2x \ d\rho \over \rho^3}
\end{equation}
 and, as $ \rho= \vert a-b\vert$ in this model is the size of instanton $z$,
this measure coincides with (\ref{measure3}).  

 In case we consider $M$ instantons the
components of the moduli space of the $2d$ instantons might be related
to the different $D(-1)-D1$ bound states.

\section{Conclusion}
\

In this paper we have analized the effects of $D$-Instantons on the
boundary of $AdS$ spaces. Though they hardly change the properties of
the space, throats joining diferent vacua are developped in their
presence. It has been also  remarked that the existence of $D-$Instantons
does not disturb the metric in the Einstein frame.

As predicted by the AdS/CFT correspondence, we have shown that these
$D$-Instantons behave as  Yang-Mills instantons, in case of $D3$-branes,
and $2d$ instantons, in case of $(D1-D5)$-branes, in the dual
pinture on the boundary and we have given exact expressions for the
corresponding gauge instantons in four and two dimensions.

We have also studied the moduli space of the supergravity solutions
finding a total correspondence to the moduli space of gauge
instantons in the case of instanton number one and  we have shown  that
the natural measure on  $AdS$ spaces is exactly the measure
of the partition function in instantonic backgrounds. We have also
discussed possible conjectures about the multiinstanton measure.

It would be interesting to go further on this subject. In particular,
a better understanding of the moduli space of $M$ $D-$Instantons  and
its correspondence to the $ADHM$ description of Yang-Mills
instantons. Another direction for future research could be the
calculation of expectation values in instantonic brackgrounds by using
 branes technology.

\section{Acknowledgements}
\

We would like to thank E. Rabinovici for useful conversations about
branes. 
The work of G.L. is supported by the spanish $FPU$ programme under grant $FP9717442117$.

\end{document}